\documentclass[aps,prb,twocolumn,superscriptaddress,showpacs,floatfix]{revtex4-1}
\usepackage{epstopdf}
\usepackage{graphicx}
\usepackage{dcolumn}
\usepackage{bm}
\usepackage{bbm}
\usepackage{color}
\usepackage{amsfonts}
\usepackage{amsmath}
\usepackage[normalem]{ulem}
\newcommand\redout{\bgroup\markoverwith{\textcolor{red}{\rule[.5ex]{2pt}{0.4pt}}}\ULon}

\usepackage[colorlinks=true,citecolor=blue]{hyperref}
\hypersetup{colorlinks=true,citecolor=blue,linkcolor=red,urlcolor=blue}

\begin{document}

\title{Valley polarized transport in strained graphene based Corbino disc}
\date{\today}

\author{Z. Khatibi}
\affiliation{School of Physics, Institute
for Research in Fundamental Sciences (IPM), Tehran 19395-5531,
Iran}
\affiliation{Physics Department, I. K. International University, PO Box 34149-16818, Qazvin, Iran }
\author{H. Rostami}
\affiliation{School of Physics, Institute
for Research in Fundamental Sciences (IPM), Tehran 19395-5531,
Iran}
\author{Reza Asgari}
\email{asgari@ipm.ir}
\affiliation{School of Physics, Institute
for Research in Fundamental Sciences (IPM), Tehran 19395-5531,
Iran}

\begin{abstract}
We study analytically and numerically the magnetotransport of strained graphene in a Corbino geometry gating in the presence of an external perpendicular magnetic field. The conductance of the Corbino disc of deformed graphene with a uniaxial and an inhomogeneous strain is calculated by using the Landauer-B\"{u}ttiker method. We show that the oscillation period of the conductance as a function of the magnetic flux depends on uniaxial strain and the conductance sharply drops along the direction of graphene stretching. The conductance amplitude, on the other hand, can be manipulated by induced pseudomagnetic flux. A valley polarized regime, caused by the inhomogeneous strain, is obtained and in addition we find a wide energy interval in which the system is fully valley polarized.

\end{abstract}

 \maketitle

\section{introduction}

Graphene has recently attracted a lot of attention as a promising
candidate material.~\cite{geim} An exciting physical feature in graphene is
strain exerted on graphene samples~\cite{vozmediano,Goerbig08,Pereira09,Ribeiro09,guinea1,diana, Rostami} and it was
proposed that strain can be utilized to generate various basic
elements for all-graphene electronics.~\cite{Pereira09}
When the graphene sheet is under external force, the side contacts
induce a long-range elastic deformation which acts as a
pseudomagnetic field for its massless charge
carriers~\cite{suzuura,maes} due to the fact that strain changes the bonds length between atoms. Its band
structure does not change for realistic strains less than
$15\%$.~\cite{Pereira09,lee08} The influence of the long-range
strains on the electronic properties is a unique feature of
graphene.~\cite{suzuura, maes} At low-energy spectrum, strains
give rise to a pseudomagnetic field which is added to the momentum
operators~\cite{vozmediano} and thus a gauge field couples to
electrons. The most evident of the unusual way in which strains
affect the electronic states is the scanning tunneling
microscope measurements of the electronic local density of states
of graphene grown on platinum.~\cite{Levy10} An average compression
of $10\%$ creates effective fields of the same order of magnitude with the value
observed in experiments.~\cite{nima} Tension can be generated either by the
electrostatic force of the underlying gate~\cite{fogler} which is caused by the interaction
of graphene with the side walls~\cite{bunch}, as a result of thermal
expansion~\cite{exp} or by quench height fluctuations.~\cite{guinea} A particular strain geometry in graphene can
lead to a uniform pseudomagnetic field and might open up interesting applications in graphene nanoelectronics
with real magnetic fields.~\cite{low} It is believed that strains have important influence on the electronic transport properties of graphene.~\cite{vozmediano}

\par
Many attempts have been made to study the bulk conductance of the Corbino geometry in two-dimensional electron gas (2DEG) systems.~\cite{boltzmann,Galdamini,Rycroft} There are experimental measurements on the charge transport of bilayer graphene~\cite{Yan10}, thermal transport~\cite{Faugeras10} and spin response of the monolayer graphene~\cite{zhao12} in the Corbino geometry.
Due to the Corbino shape, the observation of Hall effect based magnetoresistance is allowed by measuring the induced magnetic moment oscillations around the quantized value of bulk Hall conductivity.~\cite{Wiegers}
In the absence of the magnetic field, Fabry-P\'{e}rot like oscillations in the conductance of 2DEG is replaced with more moderate and suppressed ones in graphene which is a consequence of the reduced backscattering at the contacts and the absence of the details of the leads.~\cite{Rycerz09,Nazarov,Rycerz10,Katsnelson10,Rycerz12}
A periodic function of the Fermi energy which displays an insulating behavior between the Landau levels has been observed and surprisingly, oscillations independent of the magnetic flux in undoped Corbino have been also predicted theoretically.~\cite{Rycerz10}
Recently, the physical properties of graphene when its hexagonal
lattice is stretched out of equilibrium have been investigated by
many groups.~\cite{exp} Scanning tunneling microscopy
studies on the graphene surface have indeed revealed a correlation
between local strain and tunneling conductance.
\par
Besides, in valleytronics, which relies on the fact that the conduction bands of some materials have more than one minima at equal energies but at different positions in momentum space~\cite{Shkolnikov02,Beenakker,Xiao07,Isberg13}, the valley degree of freedom can be considered to produce valley polarization via controlling the number of electrons in these valleys.
Therefore, valley polarization is a key to control current and carry information analogous to spintronics. There are several proposals for generating a valley polarized current in graphene including the quantum point contact of a zigzag graphene ribbon~\cite{Beenakker}, strain~\cite{Zhai10, Chaves10, Jiang13}, introducing line defects~\cite{Gunlycke11}, helical scattering~\cite{Schomerus10} and using the effect of  trigonal warping.~\cite{PereiraJr09} Among these methods, strain is more convenient due to its intrinsic features providing the valley polarization such as introduction of a pseudogauge field with the opposite sign in the two valleys. Also there has been remarkable experimental progress in producing strain on graphene sheets.~\cite{Levy10, Klimov12} A disruptive feature which may cause great impacts on valley polarization is intervalley scattering due to the defects and imperfection of the edges which is commonplace in graphene nanoribbon and weakens the finite size system as a prosperous method for producing a valley polarized current.~\cite{Beenakker} Generally, this effect is irrelevant in the Corbino geometry. By this motivation, we would like to explore the conductance of the Corbino disc in the presence of strain.
To produce a valley polarized bulk current we consider a graphene based Corbino disc imposed on a uniaxial strain and constant pseudo and real magnetic fields. We show that, using strains both inhomogeneous and uniaxial in the absence of the edge scattering, the conductance is suppressed in one valley in such a way that the bulk conductance becomes a valley polarized in a desired direction whereas both the valleys take part in the conduction in the cross direction. We assert that by introducing strained graphene in Corbino like gating, Fig.~\ref{fig1}, one can eliminate both edge effects and extra doping. Anisotropic energy dispersion with an elliptical cross section in shifted Dirac points as a consequence of the application of a uniaxial strain has been neglected in previous studies.~\cite{Zhai10, Jiang13, low} Therefore, we would like to highlight the accuracy of the calculations based on our approach to the problem considering both shifts in the Dirac points and modifications in the band structure. We investigate the effect of strain on the oscillating nature of the conductance of the system. We also find that the oscillating period and its amplitude depend on the value and the sign of the uniaxial and inhomogeneous strain, respectively. Furthermore, we obtain the valley polarization by applying an inhomogeneous and a uniaxial strain on the Corbino disc and its dependence on the size, the value of the uniaxial strain and also the directions of the Corbino deformation.
\par
The paper is organized as follows. In Sec.~II we introduce the
formalism that will be used to calculate the electron transmission in the Corbino geometry incorporating strains and the external magnetic field. In Sec.~III
we present our analytic and numeric results for the conductance
relation and the valley polarization quantity in the system. Section~IV contains
a brief summary of our main results.
\section{Theory and Model}

We consider a strained Corbino disc in a graphene sheet in the presence of a magnetic field. It is realized that, without a magnetic field, the low-energy Hamiltonian of a uniaxially strained graphene around the shifted Dirac points can be easily described through the generalized Weyl Hamiltonian~\cite{Goerbig08,Pereira09,Rostami}, in which the uniaxial strain along a specific direction in the $(x,y)$ plane is accompanied with the modifications in the associated Fermi velocity.
\par
\begin{figure}[!htbp]
\includegraphics[width=1\linewidth]{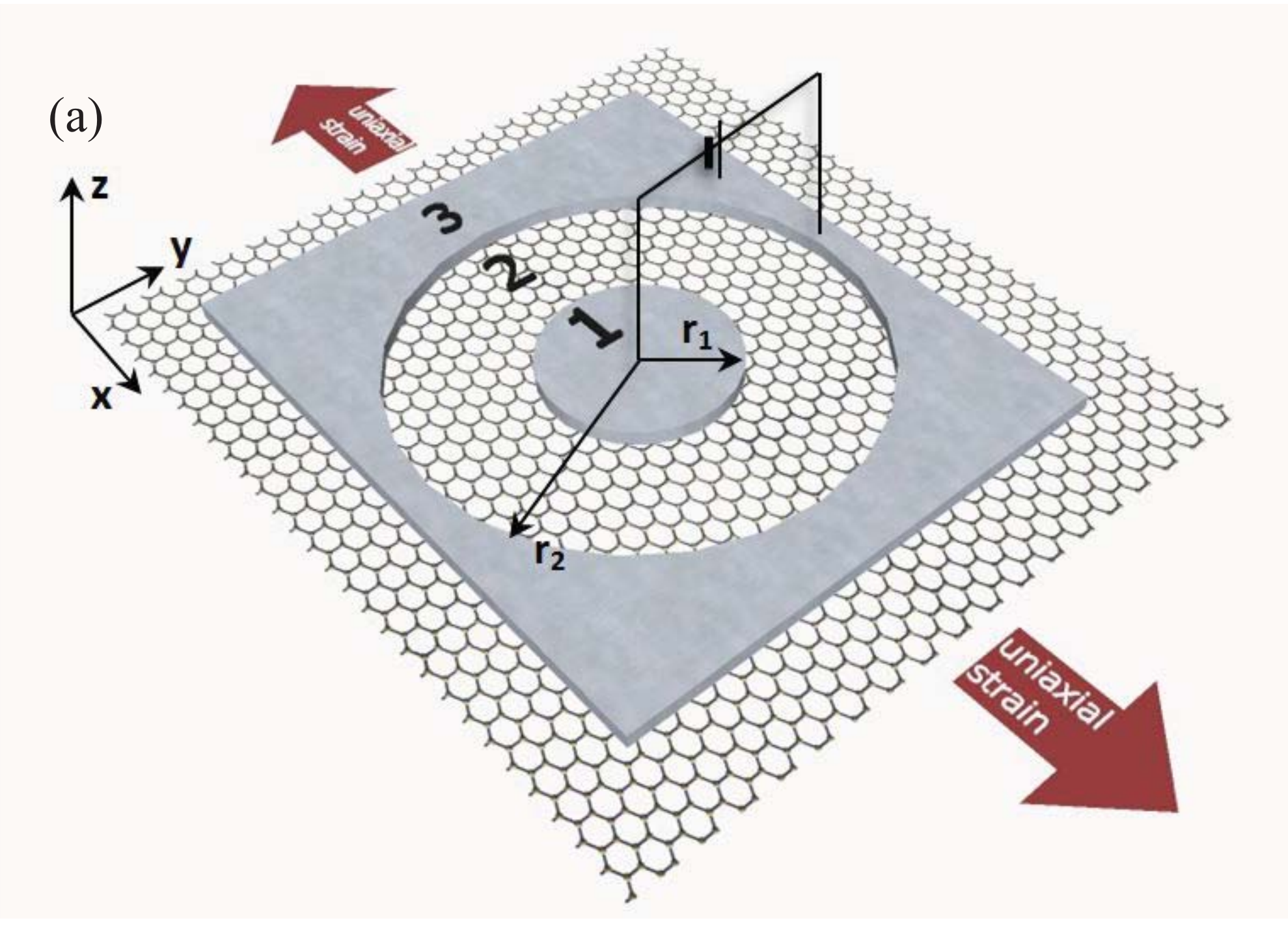}
\includegraphics[width=0.48\linewidth]{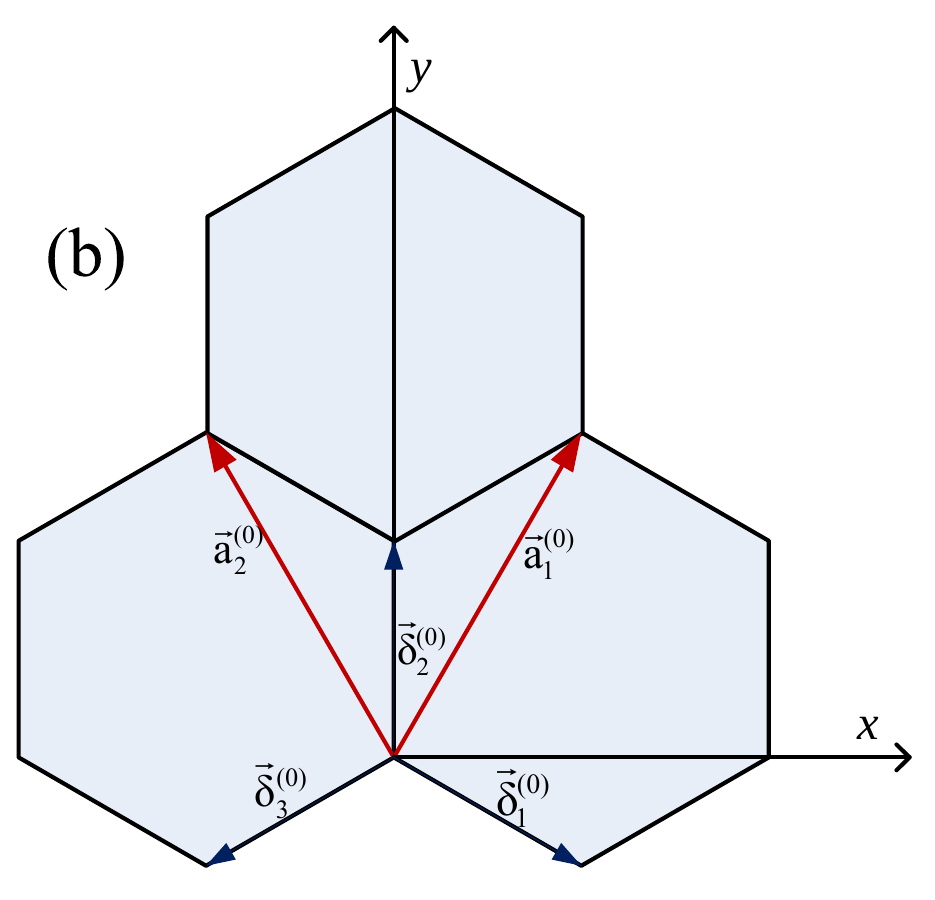}
\includegraphics[width=0.50 \linewidth]{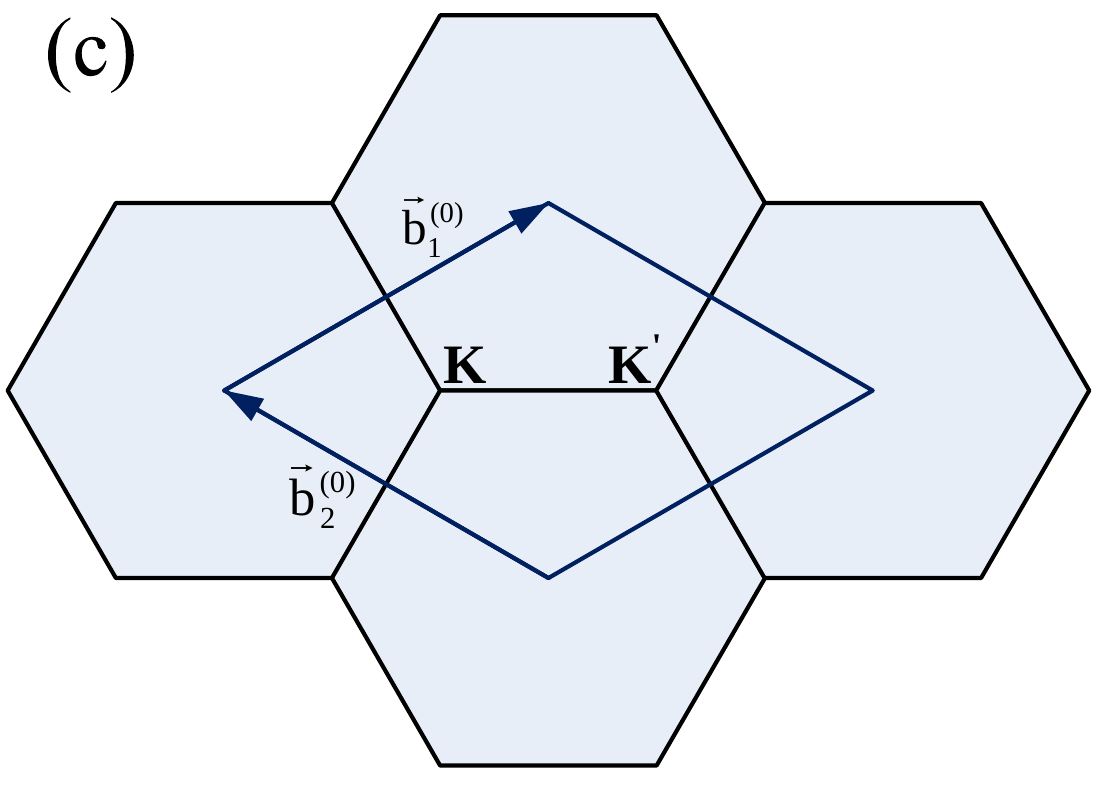}
\caption{(Color online) (a) Corbino disc when a uniaxial strain is exerted on graphene and the sample is determined by inner and outer radiuses $r_1$ and $r_2$, respectively. (b) Nearest neighbor vectors along with primitive vectors are depicted for an undeformed honeycomb lattice. (c) $K$ and $K'$ points indicate the Dirac cones location in Brillouin zone.}
\label{fig1}
\end{figure}

For the sake of completeness, we will derive the low-energy Hamiltonian incorporating the inhomogeneous strain. By applying the uniaxial strain to honeycomb structure, the hopping integrals $(t_i)$ between three nearest neighbors change from their equilibrium value, $t_0=2.7~eV$, as $t_{i}=t_0~e^{-3.37(\frac{|\vec{\delta}_{i}|}{a_0}-1)}$ (according to the previous experimental and theoretical works~\cite{Neto07,Pereira09}) in which $\vec \delta_i$ is the nearest neighbor vectors in the deformed lattice and $a_0=0.142~nm$ is $c-c$ equilibrium distance. It should be pointed out that $\vec \delta_i$ can be calculated by the nearest neighbor vectors in the undeformed case ($\vec \delta^{(0)}_i$ indicated in Fig.~\ref{fig1} (b)) and strain tensors.~\cite{Rostami} It has been shown that the uniaxial strain due to the modification of the hopping integrals creates an anisotropic energy dispersion around the new Dirac point which is shifted away from its equilibrium position in undeformed graphene. The position of the new $K$ point is~\cite{Rostami}
\begin{eqnarray}
\vec{K}_D&=&\frac{1}{2\pi}(\theta_1 \vec{b}^{(0)}_1+\theta_2 \vec{b}^{(0)}_2)\\\label{eq:K}
\theta_1 &=&\arccos{(\frac{t^2_1-t^2_2-t^2_3}{2 t_2 t_3})}\label{eq:theta1}\\ \label{eq:theta2}
\theta_2 &=&-\arccos{(\frac{t^2_3-t^2_2-t^2_1}{2 t_1 t_2})}
\end{eqnarray}
in which $b^{(0)}_{1,2}$ are the reciprocal basis vectors of the undeformed lattice (see Fig.~\ref{fig1} (c)). For a typical uniaxial strain along the $x$ direction, we have $t_1=t_3$
$\cos\theta_1=\cos\theta_2=-\frac{t_2}{2t_1}$ and $\sin\theta_1=-\sin\theta_2=\sqrt{1-(\frac{t_2}{2t_1})^2}$, therefore, the Hamiltonian close to the new Dirac point, with modified velocity along the $x $ and $y$ directions, reads
\begin{equation}
H=v_x p_x \sigma_x+v_y p_y \sigma_y
\end{equation}
which is a generalized Weyl Hamiltonian, where $v_x^2=3a_0^2(4t_1^2-t_2^2)/4\hbar^2$ and $v_y^2=9a_0^2 t_2^2/4\hbar^2$.
\par
To study the effect of pseudomagnetic field on transport properties of the system in the presence of the uniaxial strain, we apply an inhomogeneous strain to the uniaxially deformed lattice. After calculating the form factor at the new Dirac point, one can find $f(K_D)=\Sigma_i{(t_i+{\delta t}_i)e^{i\vec{K}_D.\vec{\delta}_i}}={\delta t}_1 e^{-i\vec{K}_D.\vec{a}_2}+{\delta t}_2+{\delta t}_3 e^{-i\vec{K}_D.\vec{a}_1}$, which can be written as
\begin{equation}
f(K_D)={\delta t}_1\cos\theta_2+{\delta t}_2+{\delta t}_3\cos\theta_1-i({\delta t}_1\sin\theta_2+{\delta t}_3\sin\theta_1)
\end{equation}
with $\vec{K}_D\cdot\vec{a}_i=\theta_i$. Defining a fictitious gauge field as
\begin{eqnarray}
f(K_D)&=&e v_x A^{el}_x-i e v_y A^{el}_y\\
e v_x A^{el}_x&=&{\delta t}_1\cos\theta_2+{\delta t}_2+{\delta t}_3\cos\theta_1\\
e v_y A^{el}_y&=&{\delta t}_1\sin\theta_2+{\delta t}_3\sin\theta_1
\end{eqnarray}
and substituting the value of the $\theta_1$ and $\theta_2$ in the above formulas, one can find
\begin{eqnarray}
e v_x A^{el}_x&=&{\delta t}_2-\frac{t_2}{2t_1}({\delta t}_1+{\delta t}_3)\\
e v_y A^{el}_y&=&\sqrt{1-(\frac{t_2}{2t_1})^2}({\delta t}_3-{\delta t}_1)
\end{eqnarray}
Change of the bond length values can be carried out through the method initiated in Ref.~[\onlinecite{suzuura}], where the equations of the fictitious gauge field in graphene as a function of strain tensor components ($\epsilon_{xx}$, $\epsilon_{yy}$, and $\epsilon_{xy}$), would be retrieved , $e v_x A^{el}_x= \frac{3t_2 }{4}(\epsilon_{xx}-\epsilon_{yy})$, $e v_y A^{el}_y=-\sqrt{3}\sqrt{t_1^2-t_2^2/4}~\epsilon_{xy}$.

\par
Therefore, the low-energy Hamiltonian in the presence of uniaxial strain and pseudo and real magnetic field can be generalized as
\begin{eqnarray}\label{H}
H_{\tau}&=&\tau v_x\Pi_{x}^{\tau}\sigma_{x}+v_y\Pi_{y}^{\tau}\sigma_{y}+U(r)\sigma_0,
\end{eqnarray}
with
\begin{equation}
U(r)=\left\{\begin{array}{rl}&U_0~~~~r_1<r<r_2\\ &U_\infty~~~\text{otherwise}\end{array}\right.
\end{equation}
where ${\bm \Pi^\tau}={\bf p}+e{\bf A}+e \tau {\bf A}^{el}$ with ${\bf A}$ denoting the real magnetic field and ${\bf A}^{el}$ standing for the pseudogauge field. $\tau (\pm )$ indicates the Dirac point $(K,K^\prime)$.
Notice that we have neglected the material independent term because that term doesn't have any contribution to the pseudomagnetic field.~\cite{Fernando13}

We are interested in application of a symmetric gauge ${\bf A}=\frac{B}{2}(-y,x)$ considering the Corbino geometry. With the choice of an appropriate inhomogeneous deformation added to the uniaxially deformed graphene, one can induce a pseudogauge field as ${\bf A}^{el}=(\lambda_x A_x,\lambda_y A_y)$ which is proportional to the real gauge field. A symmetric fictitious gauge field, by choosing the preferred strain profile as $u_x=-2u_0 xy/a_0$ and $u_y=u_0(y^2-x^2)/a_0$, is given through
\begin{eqnarray}
\lambda_x&=&\frac{6u_0t_2}{Bev_x a_0}\nonumber \\
\lambda_y&=&4 \sqrt{3}\frac{u_0 \sqrt{t_1^2-t_2^2/4}}{Bev_y a_0}
\end{eqnarray}
where $u_0$ is a dimensionless parameter indicating the magnitude of deformation and, for example equals to $u_0=7.6\times 10^{-6}$ for $B^{el}=1T$ at $\epsilon=0$.

Notice that we assume that the circular leads are highly doped ($U_\infty\gg U_0$). Consequently, we neglect the effect of the real magnetic and pseudo-magnetic fields on these regions, since the high-energy electrons in heavily doped leads are immune to the Landau levels introduced by strain and the real magnetic field. In other words, high-energy electrons can not be trapped by magnetic field with reasonable strength and hence the effect becomes ignorable on them.
The eigenvalue problem of the Hamiltonian leads to a second order differential equation for each pseudospin component, and should be solved in the three regions of the Corbino system denoted in Fig.~\ref{fig1}. The main eigenvalue problem leads to
\begin{equation}\label{eigen}
[v^2_x\Pi^{\tau2}_x+v^2_y \Pi^{\tau2}_y+l_{ps} eB \hbar(\lambda+\tau )v_xv_y-(E-U(r))^2 ]\Psi^{\tau l}(x,y)=0
\end{equation}
in which $l=\uparrow (\downarrow)$ denotes the lattice pseudospin with corresponding $l_{ps}=+(-)$ and $\lambda=(\lambda_x+\lambda_y)/2$. Further simplifying Eq.~(\ref{eigen}), a second order differential equation can then be achieved
\begin{widetext}
\begin{eqnarray}\label{psiud}
\nonumber
[v^2_x p^2_x+v^2_y p^2_y+\tau eB(\lambda+\tau )(v_y^2p_yx-v_x^2p_xy)+(\frac{eB}{2})^2(\lambda+\tau )^2((yv_x)^2+(xv_y)^2)\\
l_{ps} eB \hbar(\lambda+\tau )v_xv_y] \Psi^{\tau l}(x,y)=(E-U(r))^2\Psi^{\tau l}(x,y)
\end{eqnarray}
\end{widetext}
We neglect the term $\Delta\lambda~ {\mathcal D}(x,y)\Psi^{\tau l}(x,y)+\mathcal{O}(\Delta\lambda^2)$ in deriving Eq.~(\ref{psiud}), where
\begin{eqnarray*}
{\mathcal D}&=&(\frac{eB}{2})^2( \lambda+\tau)[(yv_x)^2-(xv_y)^2]\\
&-& \tau \frac{eB}{2}[y v_x^2p_x+x v_y^2p_y]
\end{eqnarray*}
since strained hopping energies and velocities along the $x$ and $y$ directions up to the first order of the $\epsilon$ in a uniaxially (i.e. $\epsilon_{xx}=\epsilon$ , $\epsilon_{yy}=-\nu \epsilon$, and $\epsilon_{xy}=0$) deformed  graphene are
\begin{eqnarray*}
v_x&\approx& v_{\rm F}(1-3.37 \epsilon)\\
v_y&\approx&v_{\rm F}(1+3.37 \epsilon \nu)\\
t_2&\approx &1+3.37\nu \epsilon\\
t_1=t_3&\approx &1-\frac{3.37}{4}(3-\nu) \epsilon,
\end{eqnarray*}
where $\nu=0.165$ is the Poisson's ratio and $v_{\rm F}\simeq 10^6 m/s$ is the Fermi velocity. Therefore, the term $\lambda_x-\lambda_y=u_0 \epsilon 40.44 (1+\nu)/(Bev_{\rm F}a_0)$ is ignorable, since it is proportional to the second order of strain. We will first take the steps to solve the equation of the down component of the pseudospin and then find the other one with help of the first order coupled equation which is given by
\begin{equation}\label{psiu}
\Psi^{\tau\uparrow}(x,y)=\frac{1}{E-U(r)}(\tau v_x\Pi_{x}^{\tau}-iv_y\Pi_{y}^{\tau})\Psi^{\tau \downarrow}(x,y)
\end{equation}
We transform the equations into a new coordinates, using new variables ($R$, $\Theta$), where boundary conditions can be applicable. A straightforward calculation on Eq.~(\ref{psiud}) yields
\begin{equation}\label{psid}
[\partial^2_R+\frac{1}{R}\partial_R+\frac{\partial^2_\Theta}{R^2}+\tilde E^2+2\gamma(i\tau \partial_\Theta-l_{ps})-\gamma^2R^2]\Psi^{\tau l}(R,\Theta)=0
\end{equation}
where we have introduced the following dimensionless variables
\begin{eqnarray}
R(r,\theta)&=&\frac{r}{v_{\rm F}r_s}\sqrt{v^2_y\cos^2{\theta}+v^2_x\sin^2{\theta}}\nonumber\\
\Theta(r,\theta)&=&\tan^{-1}{(\frac{v_x\sin{\theta}}{v_y\cos{\theta}})}\nonumber\\
\gamma&=&\frac{e\tilde Br^2_s}{2 \hbar},~~~~\tilde E=\frac{E-U(r)}{E_s}
\end{eqnarray}
in which $\tilde B=B( \lambda+\tau)~v^2_{\rm F}/v_xv_y$ and $E_s=\hbar v_xv_y/v_{\rm F}r_s$. A typical length scale of the system, $r_s$, which is considered to be equal to the inner radius, is introduced. For more simplification in notations, hereafter, we drop the $r$ and $\theta$ dependence of $R(r,\theta)$ and $\Theta(r,\theta)$.

Since operator $\hat{\cal{O}}=XP_Y-YP_X+\sigma_z/2$, in which $(X,Y)=v_{\rm F }r_s(R\cos\Theta,R\sin\Theta)$, commutes with the Hamiltonian, its eignestate, $e^{im\Theta}$, is the eignestate of the Hamiltonian simultaneously where $m$ is the associated quantum number with an integer value.
By substituting $\Psi^{\tau\downarrow}=e^{i(m+1)\tau \Theta}\Phi^{\tau \downarrow}(R)$ in Eq.~(\ref{psid}), the differential equation reduces to the confluent hypergeometric. The $R$-dependent term of the wave function reads to the Hankel function in the absence of the real and pseudomagnetic fields in the inner and the outer leads $(r<r_1,r>r_2)$. Eventually, the Corbino states in the inner and the outer leads become
\begin{eqnarray}\label{psi1}
\nonumber
\Psi^\tau_{1}(R,\Theta)=e^{im\tau \Theta}[\begin{pmatrix} -i~\rm{sgn} (\tilde E_\infty)H^1_{m}(|\tilde E_\infty|R)\\H^1_{m+1}(|\tilde E_\infty|R)~e^{i\tau \Theta} \end{pmatrix}\\
+\kappa\begin{pmatrix} -i~\rm{sgn}(\tilde E_\infty)H^2_{m}(|\tilde E_\infty|R)\\H^2_{m+1}(|\tilde E_\infty|R)~e^{i\tau \Theta} \end{pmatrix}]
\end{eqnarray}
\begin{equation}\label{psi3}
\Psi^\tau_{3}(R,\Theta)=t~e^{im\tau \Theta} \begin{pmatrix} -i~\rm{sgn} (\tilde E_\infty)H^1_{m}(|\tilde E_\infty|R)\\H^1_{m+1}(|\tilde E_\infty|R)~e^{i\tau \Theta} \end{pmatrix}
\end{equation}
where $\kappa$ and $t$ are the reflection and the transmission amplitudes, respectively.
Moreover, the corresponding wave function between the two leads is given by
\begin{align}\label{psi2}
\nonumber
\Psi^\tau_{2}(R,\Theta)=e^{im\tau \Theta} R^{|{m+1}|}&e^{-\frac{|\gamma|R^2}{2}}\\
&\begin{pmatrix}-i(A~M^\uparrow(R)+C~U^\uparrow(R))\\(A~M^\downarrow(R)+C ~U^\downarrow(R))~e^{i\tau \Theta}\end{pmatrix}
\end{align}
where
\begin{eqnarray}
Z^{\sigma}(R)&=&(\delta_{\sigma\uparrow}f(R)/\tilde{E_0}+\delta_{\sigma\downarrow})Z(\alpha,\beta,|\gamma|R^2)\nonumber\\
&+&(\delta_{\sigma\uparrow}/\tilde{E_0}) (2\gamma R\alpha/\xi_Z) Z(\alpha+1,\beta+1,|\gamma|R^2)\nonumber \\
\end{eqnarray}

in which $Z(x,y,z)$ can be either $M(x,y,z)$ or $U(x,y,z)$ denoting the confluent hypergeometric functions~\cite{mathfunc} with $\alpha=\beta/2-(\tilde E_0^2-2\gamma m)/4|\gamma|$ and $\beta=|m+1|+1$. We have $\xi_M=\beta$, and $\xi_U=-1$ for the two kinds of the hypergeometric functions. Furthermore, $f(R)=(\gamma-|\gamma|)R+(m+1+|m+1|)/R$, $\delta_{ij}$ is the Kronecker delta and
$\tilde E_{0,\infty}=(E-U_{0,\infty})/E_s$ is namely the doping rate. $A$ and $C$ are also two constants that can be determined through boundary conditions.

We solve matching boundary conditions to find the transmission probability. Since the $\Theta$ dependence of the wave function is ineffective, the procedure in both valleys therefore reduces to only the $\Phi (R)$ continuity constraint as
\begin{eqnarray}
\Phi_1^{\tau l}(R(r_1,\theta))&=&\Phi_2^{\tau l}(R(r_1,\theta))\label{b1}\nonumber\\
\Phi_2^{\tau l}(R(r_2,\theta))&=&\Phi_3^{\tau l}(R(r_2,\theta ))\label{b2}
\end{eqnarray}
and the electron transmission is given by $T_m(E,\theta, \tau)=|t|^2$. We anticipate a direct dependence of the transmission probability on $\theta$ since the uniaxial strain results in anisotropic dispersion relation.
Using $H_m^1(\rho)=\sqrt{\frac{2}{\pi \rho}}~\exp{(i(\rho - m \pi /2-\pi /4))}$ as the asymptotic behavior of the Hankel functions in highly doped leads, the transmission probability is thus given by
\begin{eqnarray}\label{trans}
T_m(E,\theta,\tau)=(\frac{r_2}{r_1})^{2\beta-1}
\frac{4e^{|\gamma|(R^2_1-R^2_2)}~{\eta^{\downarrow\uparrow}_{22}}^2}{(\eta^{\uparrow\uparrow}_{12}-\eta^{\downarrow\downarrow}_{21})^2+
(\eta^{\uparrow\downarrow}_{12}-\eta^{\downarrow\uparrow}_{21})^2}
\end{eqnarray}
where $\eta^{\sigma\sigma'}_{ij}=M^{\sigma}(R_i)U^{\sigma'}(R_j)+\sigma.\sigma'M^{\bar{\sigma}}(R_i)U^{\bar{\sigma'}}(R_j)$ and $R_{i:1,2}$ indicates $R(r_i,\theta)$.

To evaluate the disc conductance, we then use the Landauer-B\"{u}ttiker formula which presumes conducting through the channels of different modes and sums up over all modes, considering all channels contribution to the conductance of a two terminal system in zero temperature:
\begin{equation}\label{g}
G_{\tau}=\frac{2 e^2}{h}\sum_m{T_m(E,\theta,\tau)}
\end{equation}
Here the factor $2$ stands for spin degeneracy. Notice that the conductance, in the Corbino geometry, is attainable from bulk channels which means proportionality to $\sigma_{xx}$.

For better understanding, we consider the conductance in the zero doped Corbino disc, where setting $\tilde{E_0}=0$, causes reduction in the hypergeometeric functions~\cite{mathfunc} as follows\\
case $m+1>0$:
\begin{eqnarray}\label{casep}
\nonumber
\alpha_+ &=& m+1~~~,~~~\beta_+ = m+2,\\ \nonumber
M_i^\uparrow &=&\frac{2(m+1)~e^{c_i}}{\tilde{E_0}R_i}\\ \nonumber
M_i^\downarrow &=&(m+1)(-c_i)^{-(m+1)}\gamma(m+1,-c_i)\\
U_i^\uparrow &=&0~~~,~~~U_i^\downarrow =c_i^{-(m+1)}
\end{eqnarray}
case $m+1<0$
\begin{eqnarray}\label{casem}
\nonumber
\alpha_- &=& 0~~~,~~~\beta_- = -m,\\ \nonumber
M_i^\uparrow &=&\frac{-2~c_i~\alpha_-}{m~\tilde{E_0}R_i}~e^{c_i}~c_i^m(\Gamma(1-m)+m\Gamma(-m,c_i))\\ \nonumber
U_i^\uparrow &=&\frac{-2~c_i~\alpha_-}{~\tilde{E_0}R_i}~e^{c_i}~c_i^m\Gamma(-m,c_i),\\
M_i^\downarrow &=& 1~~~,~~~U_i^\downarrow=1
\end{eqnarray}
where we consider the real magnetic field in the $z$ direction, i.e. $|\gamma|=\gamma$ and $c_i=\gamma R_i^2$.

By substituting the new format of the Hypergeometric functions Eqs.~(\ref{casep}), (\ref{casem}) in Eq.~(\ref{trans}), the transmission probability in both cases reduces to
\begin{eqnarray}\label{trans0}
T_m(\tilde{E_0}\rightarrow0,\theta,\tau)=\frac{1}{\rm{\cosh^2}((m+1/2)\mathcal{L+X})}
\end{eqnarray}
with $\mathcal{L}=\rm{ln}(R_2/R_1)$ and $\mathcal{X}=\gamma(R^2(r_2,\theta)-R^2(r_1,\theta))/2$. Then after summing over the modes~\cite{Rycerz10}, the disc conductance is equal to
\begin{equation}
\label{gsdisc}
G=g_0\sum_{m}T_m(\tilde{E_0}\!\rightarrow\!{0},\theta,\tau)=
\sum_{j=0}^\infty G_j\cos\left(\frac{2\pi{j}\mathcal{X}}{\mathcal{L}}\right),
\end{equation}
where $g_0=4e^2/h$ is the quantum conductance and the Fourier amplitudes are
\begin{equation}\label{G-asymptotic}
G_0=\frac{2g_0}{\mathcal{L}}, \ \ \ \ \
G_{j}=\frac{4\pi^2(-)^jjg_0}{\mathcal{L}^2\sinh(\pi^2j/\mathcal{L})}
\ \ \ (j\!>\!{0}).
\end{equation}
in which the effect of strain appears in the conductance through $\mathcal{L}$ and $\mathcal{X}$. Importantly, we can learn from this equation, which is our main analytic expression, that the oscillation amplitudes are independent of the applied uniaxial strain whereas the oscillation periods are strongly affected by the direction and value of strain.

Since the pseudomagnetic field differs in sign in the two valleys, the conductance of an electron is therefore different in each valley, which causes manifestation of the valley polarization of the longitudinal conductance in the Corbino geometry:
\begin{equation}\label{pol}
P=\frac{G_K-G_{K^\prime}}{G_K+G_{K^\prime}}
\end{equation}
In a standard ribbon system the valley polarization, due to the strong intervalley scattering at the edges, is suppressed by the pseudomagnetic field~\cite{low} and the finite size effect~\cite{Beenakker} whereas in the Corbino geometry, this effect is irrelevant, due to the absence of the edges. Furthermore, in spite of the standard valley Hall polarization which comes from the opposite responses of the different valleys in the Hall system, $\sigma^K_{xy}-\sigma^{K^\prime}_{xy}$, it is proportional to $\sigma^K_{xx}-\sigma^{K^\prime}_{xx}$ in the Corbino case.

\section{Numerical Results}
In this section, we present our obtained numerical results for the magnetotransport of a strained graphene in the Corbino gating in the presence of an external perpendicular magnetic field. Using the Landauer-B\"{u}ttiker method, the conductance of the Corbino disc in deformed graphene geometry with a uniaxial and an inhomogeneous strain based on Eqs.~(\ref{psi1}), (\ref{psi3}, (\ref{psi2}) and (\ref{g}) is calculated and the results are compared with those results for an unperturbed disc. A valley polarized regime, caused by inhomogeneous strain is observed as well. We therefore provide the corresponding valley polarization results based on Eq.~(\ref{pol}). In all figures we assume $t_0=2.7$eV, $r_1=10$nm as the nearest neighbor hopping integral and the inner lead radius, respectively.

\begin{figure}[!htbp]
\centering
\includegraphics[width=\linewidth]{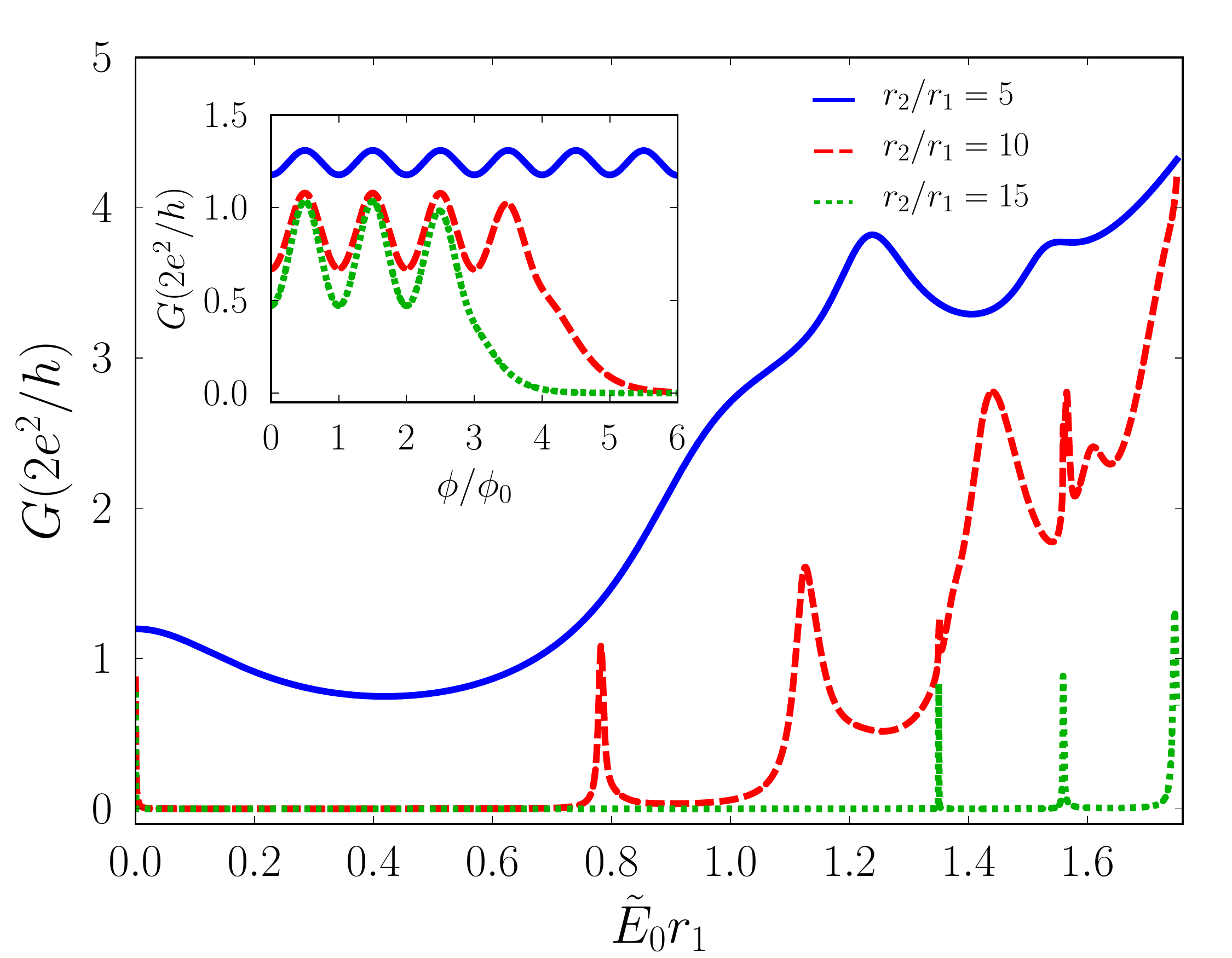}
\caption{(Color online) Conductance as a function of doping for diverse radius ratios and B=2T. Pseudodiffusive, field suppressed and ballistic regimes are clearly observed. For the small radius ratio, the magnetic field cannot localize the particle, consequently the system is in a ballistic phase. By increasing the size of the scattering region, the field suppression region increases. Inset: Illustration of the size dependence of the Corbino oscillations for $\tilde E_0r_1=10^{-4}$ and $\phi_0=2(h/e)\mathrm{ln}(r_2/r_1)$.~\cite{Rycerz10}}
\label{fig2}
\end{figure}

As it was previously shown~\cite{Rycerz09}, quantization steps in the conductance which were first observed in graphene strips, are absent in the Corbino disc. Moreover, the conductance shows a weak oscillation rather than a linear dependence of the doping rate, as a consequence of the lack of the back scattering in comparison with 2DEG systems. In the presence of the magnetic field the linear dispersion of graphene changes to the flat bands regime of the Landau levels with a wide gaps among them. In this case, we find ballistic pseudodiffusive and field suppressed regimes~\cite{beenakker} in the phase space of the size and the magnetic field strength.~\cite{Prada07} In the case of $ 2l_B^2k_{\rm F}/(r_2-r_1)<1$,
in which $l_B$ and $k_{\rm F}$ are magnetic length and the Fermi wave vector respectively, the system is no longer in a ballistic regime and it displays a crossover from the pseudodiffusive regime, close to the Landau levels, to the field suppressed regime when one stands far from the Landau levels. Consistent with previous works the results are shown in Fig.~\ref{fig2} for different sizes of the system. We should notice that this oscillation in the conductance as a function of magnetic flux is due to the Fabry-Perot interference which is very similar to the quantum interference observed in a two-barrier system for non-relativistic carriers.~\cite{Nazarov} However, the moderate oscillation in the graphene case is due to Klein tunneling mechanism for its massless Dirac particles. This effect has been also addressed by some previous theoretical studies.~\cite{Katsnelson10,Rycerz12}

It is clear that the size of the system has a dramatic effect on the conductance and a stronger field suppression appears in a larger Corbino which can be helpful to generate the valley polarized current as we will discuss later. Formation of the Landau levels restricts the conduction into the Landau channels whose energies are proportional to $\sqrt{n B}$. It is then trivial that as a consequence of the increase in the magnetic flux piercing the disc area, the number of the channels contributing to the conduction reduces till all Landau levels (except zero level) stand beyond the doping rate of the disc area. As a result of the Corbino geometry an oscillating behavior in the conductance for different sizes shows up as it is illustrated in the inset of Fig.~\ref{fig2}. The increase in the size of the Corbino disc modifies the oscillation period and the amplitude as indicated in the inset.
\begin{figure}[!htbp]
\includegraphics[width=\linewidth]{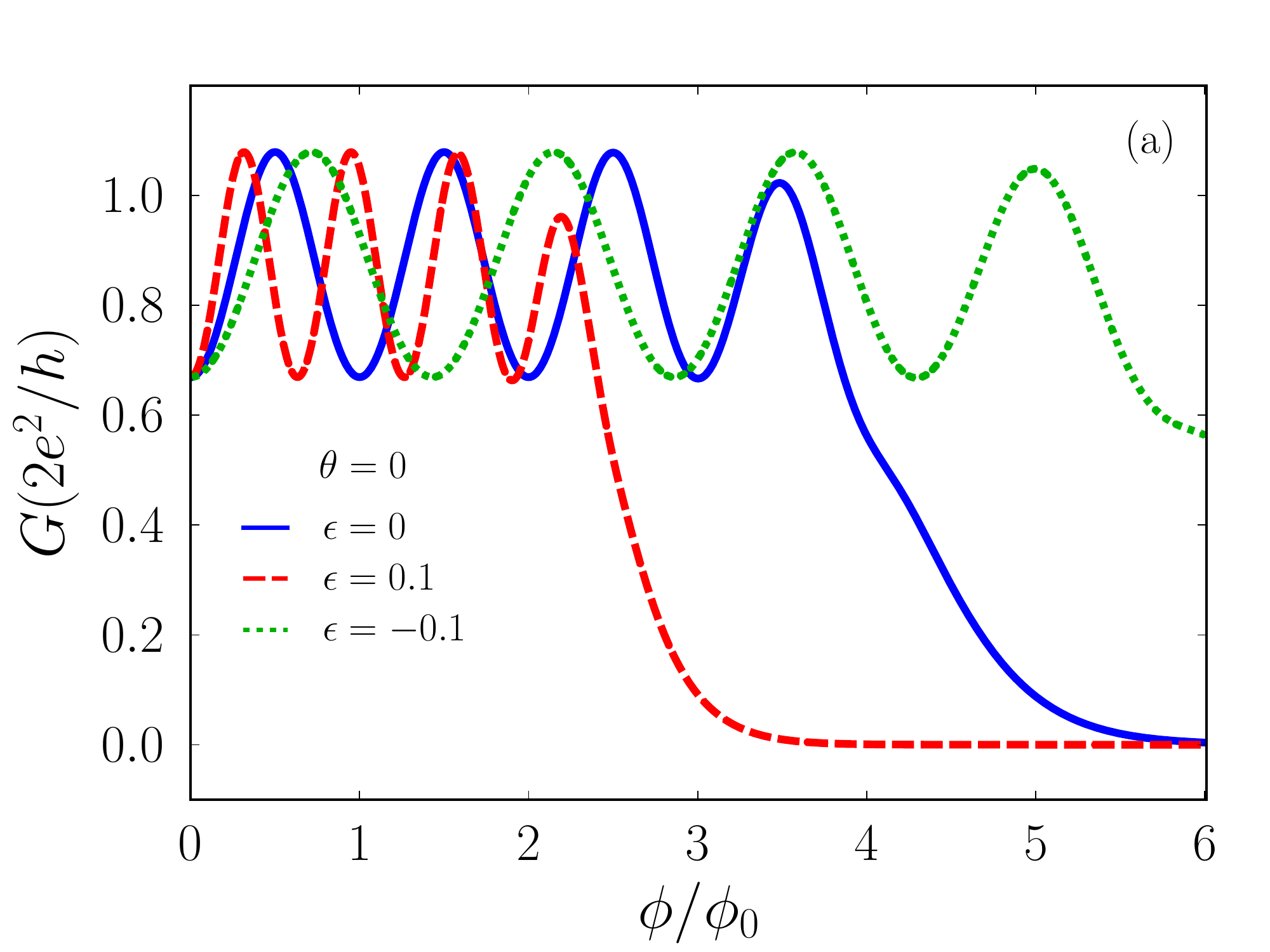}
\includegraphics[width=\linewidth]{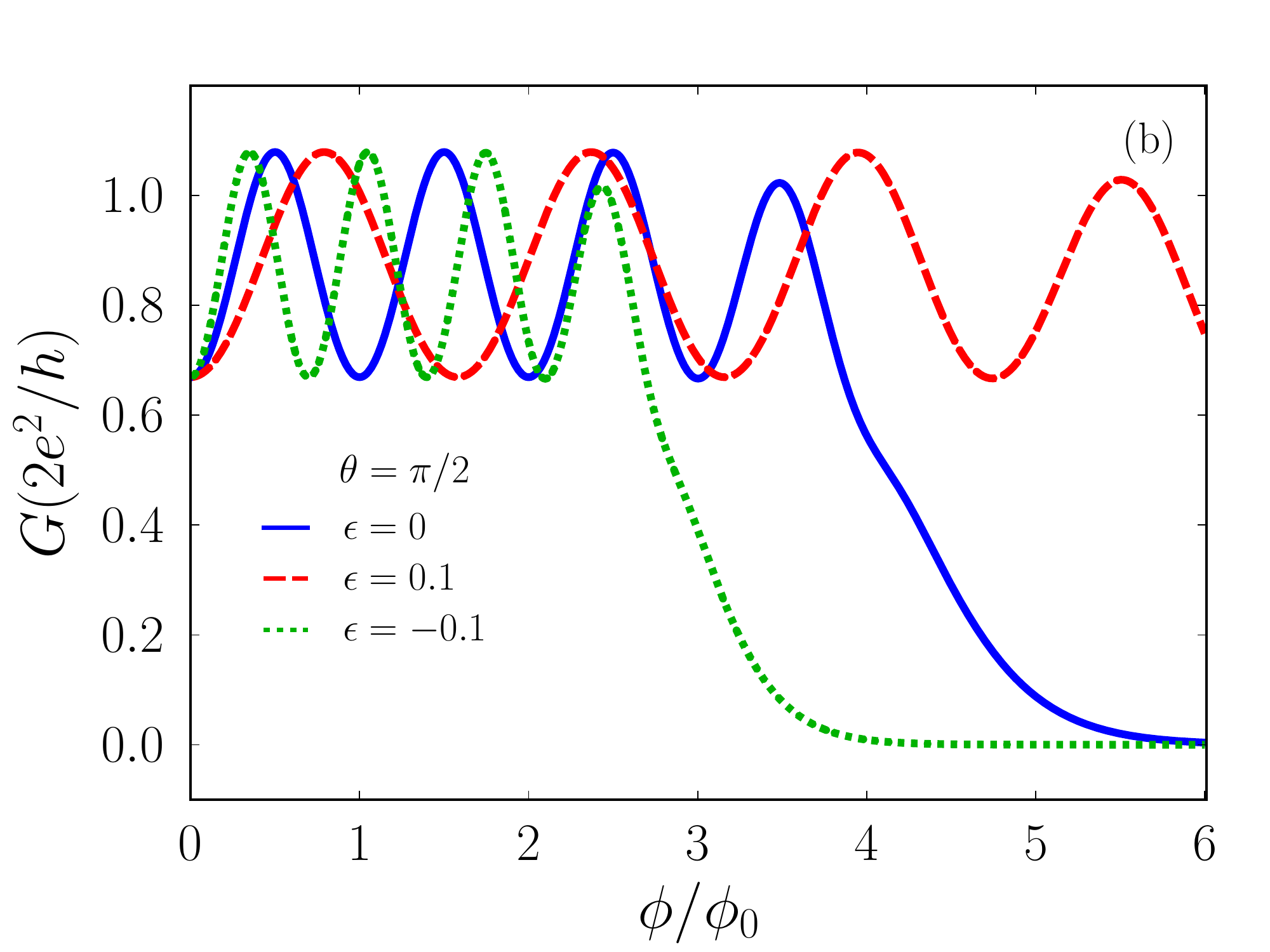}
\includegraphics[width=\linewidth]{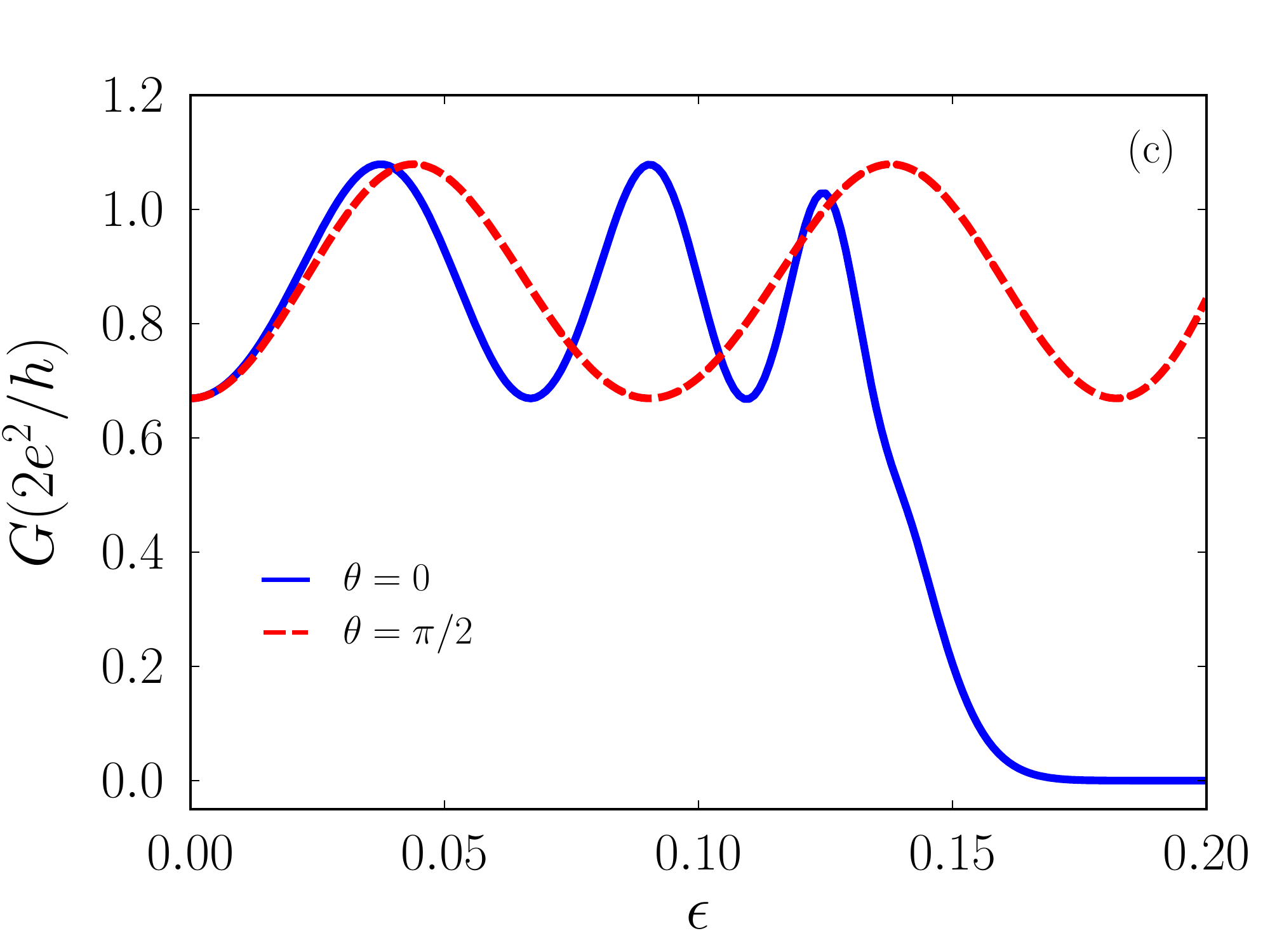}
\caption{(Color online) Conductance oscillations for different magnitudes of the uniaxial strain. (a) and (b): strain tuning effect on the period of the oscillations along the $x$ and the $y$ directions are shown, respectively for the $\tilde{E}_0r_1=10^{-4}$. (c) demonstrates a non-trivial oscillation in the conductance as a function of strain for $\tilde{E}_0r_1=10^{-6}$, $r_2=10r_1$ and $\phi/\phi_0=3$.}
\label{fig3}
\end{figure}

Now, we investigate the effect of the uniaxial strain on the oscillating nature of the conductance as a function of the magnetic flux along the $x$ and $y$ directions which is demonstrated in Fig.~\ref{fig3}(a) and Fig.~\ref{fig3}(b), respectively. Due to the anisotropic dispersion of the uniaxially strained graphene, the charge velocities along the $x$ and $y$ directions differ from each other. This is why one witnesses earlier charge confinement in the $x$ direction and a delay in the $y$ direction rather than the case without the application of the uniaxial strain. A delightful point is that the uniaxial strain does not change the oscillation amplitude but its period, which is better manifested in the figures. In Fig.~\ref{fig3}(c) the conductance as a function of strain is brought, in which a non-trivial oscillation is clear with a sharp decay along the $x$ direction. These analysis are in good agreement with the asymptotic expressions of the conductance given by Eq.~(\ref{G-asymptotic}).

\begin{figure}[!htbp]
\includegraphics[width=\linewidth]{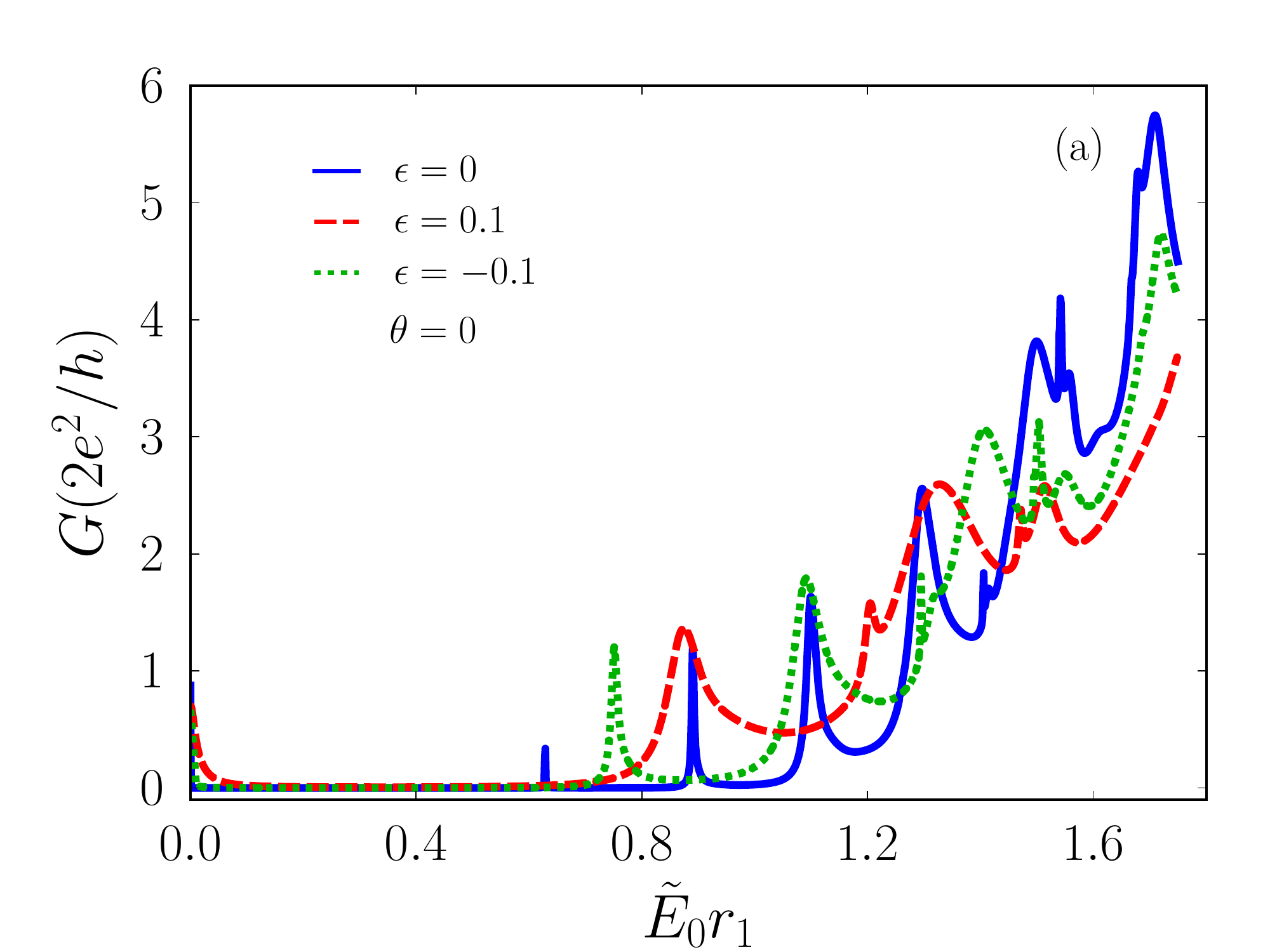}
\includegraphics[width=\linewidth]{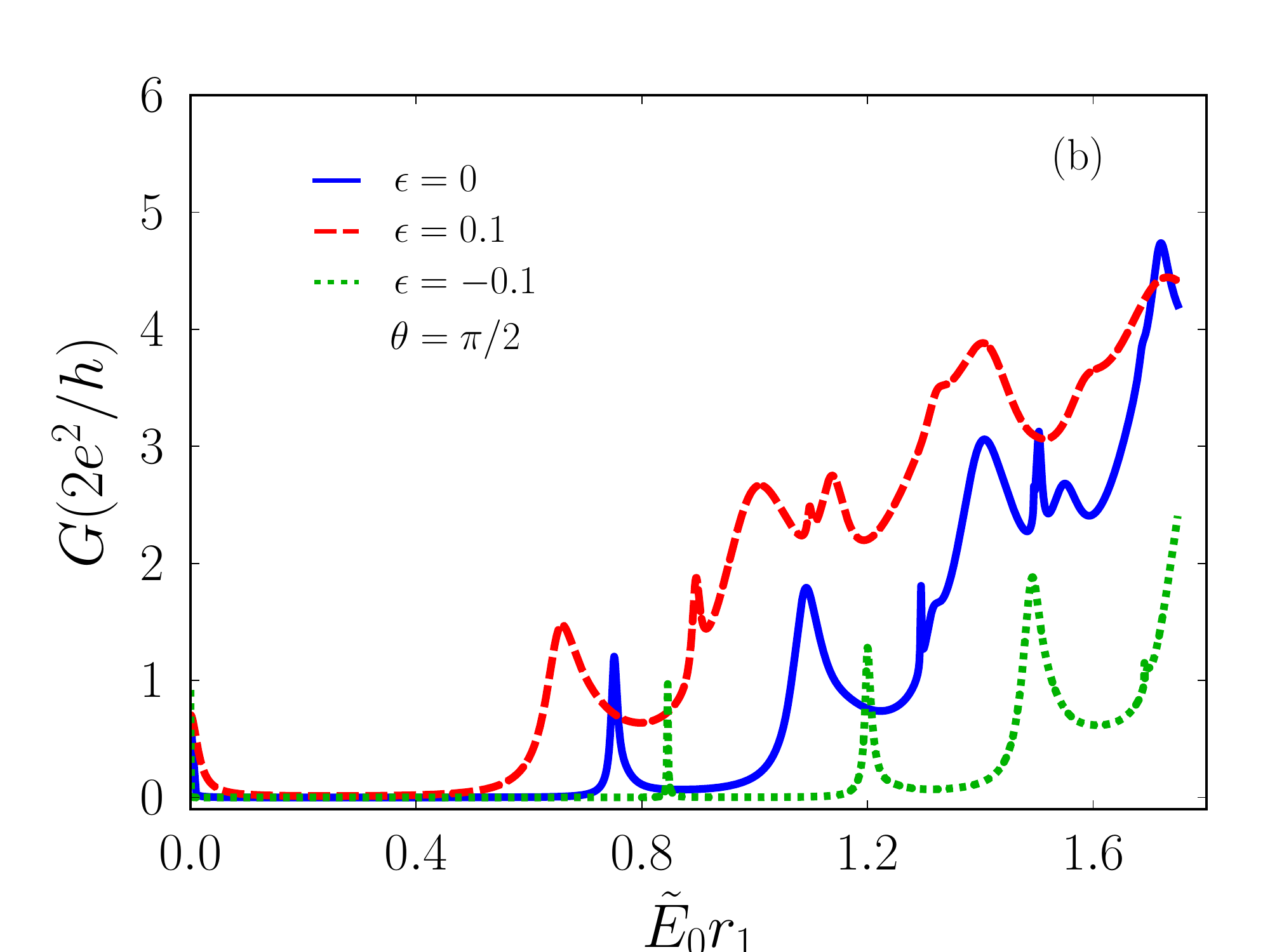}
\caption{(Color online) Conductance as a function of doping for different values of strain: (a) and (b) correspond to the $x$ and the $y$ directions, respectively where strain makes a shift in the position of the resonances. $\phi=3\phi_0$, $r_2=10r_1$ and ${\bf A}^{el}=0$.}
\label{fig4}
\end{figure}

Figure~\ref{fig4} indicates the strain dependence of the conductance as a function of doping in the $x$ and the $y$ directions, from which one can find a strong modulation of the field suppression region and the position and height of the peak in conductance.
In the uniaxially strained graphene, the conductance in the valleys (shifted Dirac points) is the same despite the fact that the time reversal symmetry is broken due to the non-zero perpendicular magnetic field. There is no valley polarized current which is inconsistent with the the valley polarized conductance in the uniaxially strained graphene strip.~\cite{Zhai12,Zhai10} However one can generate the valley polarized bulk current introducing an inhomogeneous strain which induces a constant pseudomagnetic field (${\bf B}^{el}$) with opposite sign in the two valleys, in such a way that electrons in the valleys feel different total magnetic fields, ${\bf B}\pm{\bf B}^{el}$. We consider ${\bf B}^{el}=\lambda{\bf B}$ in which $\lambda$ depends on a homogenous and inhomogeneous strain.

\begin{figure}[!htbp]
\includegraphics[width=\linewidth]{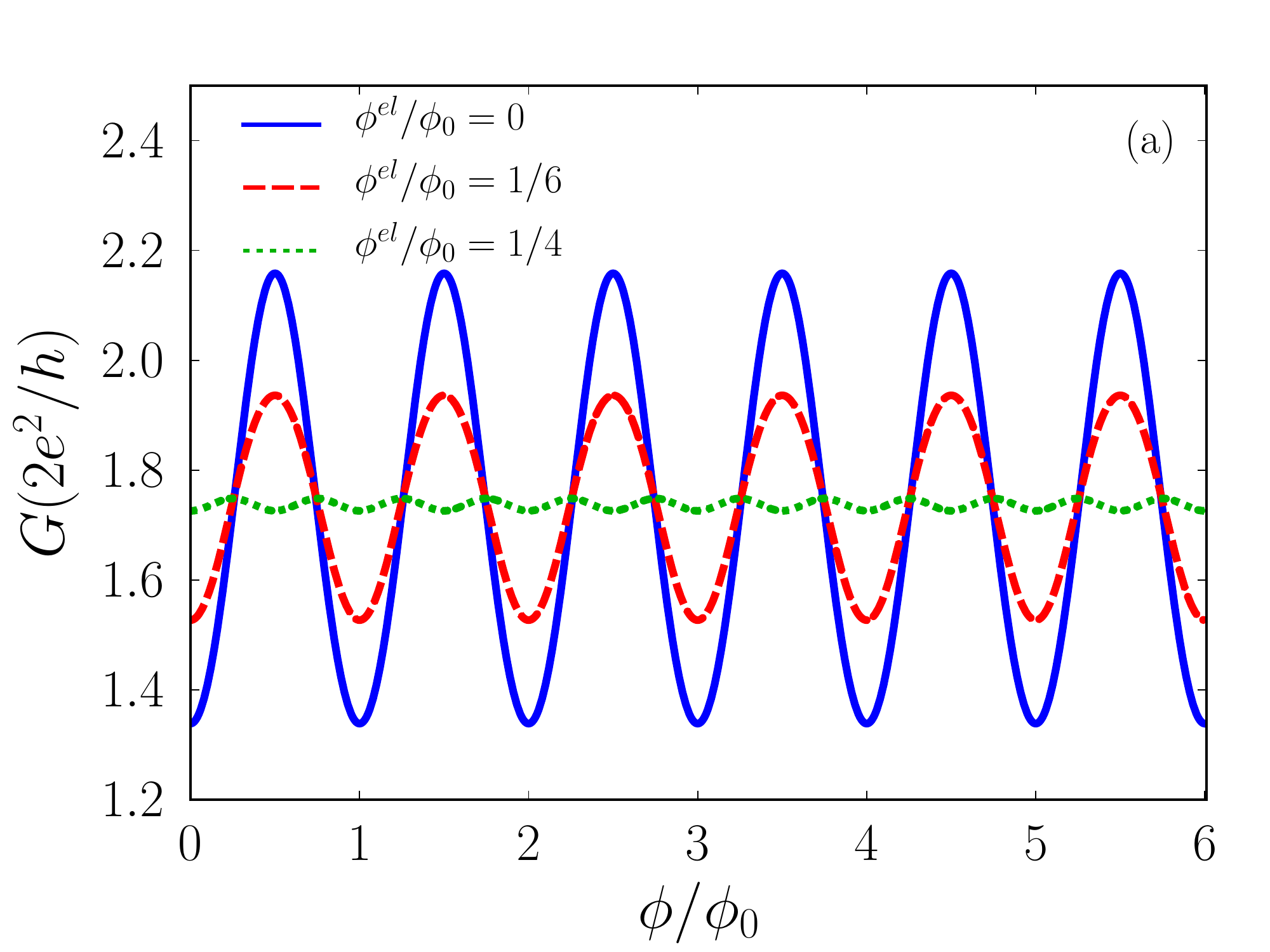}
\includegraphics[width=\linewidth]{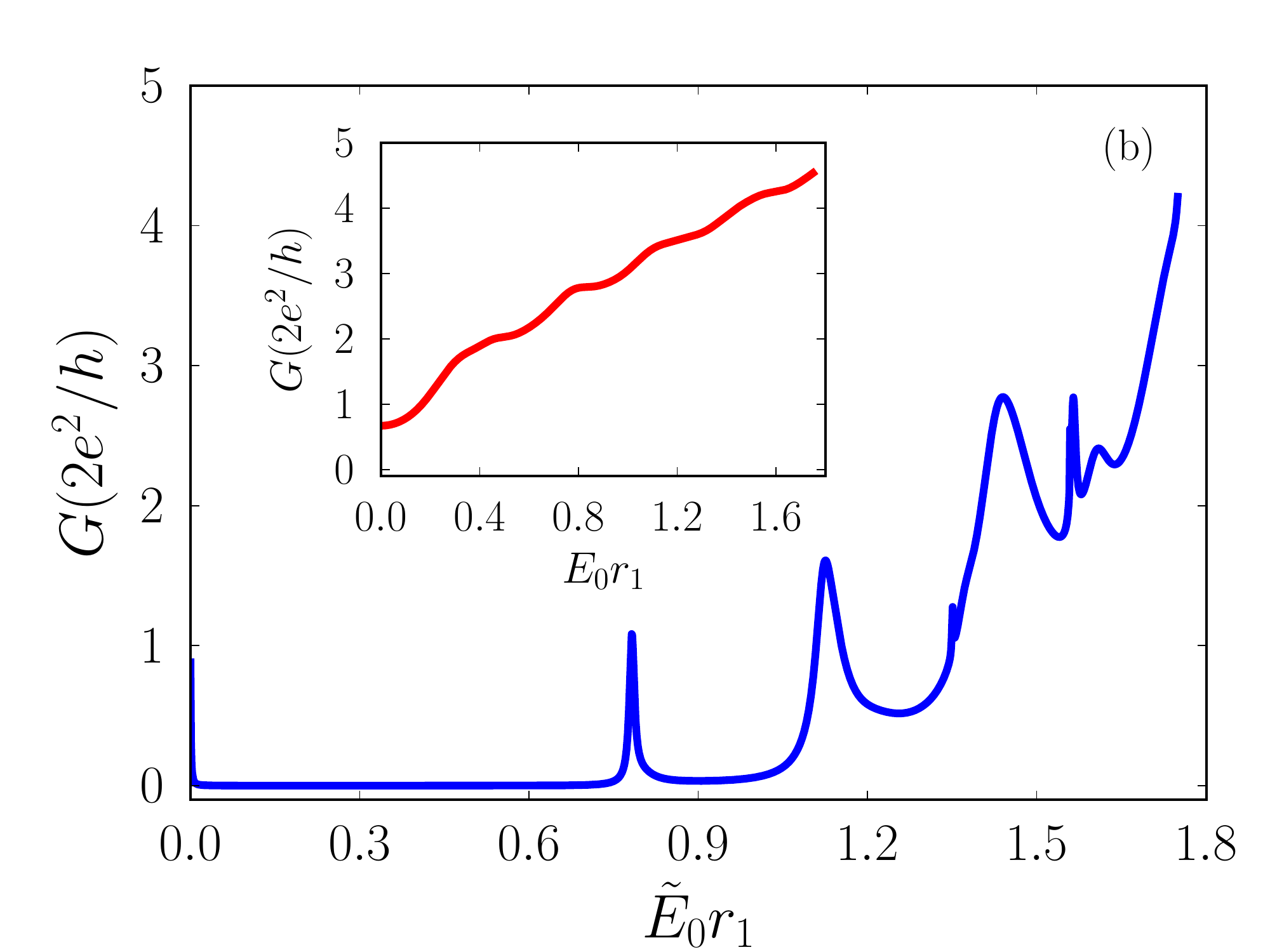}
\includegraphics[width=\linewidth]{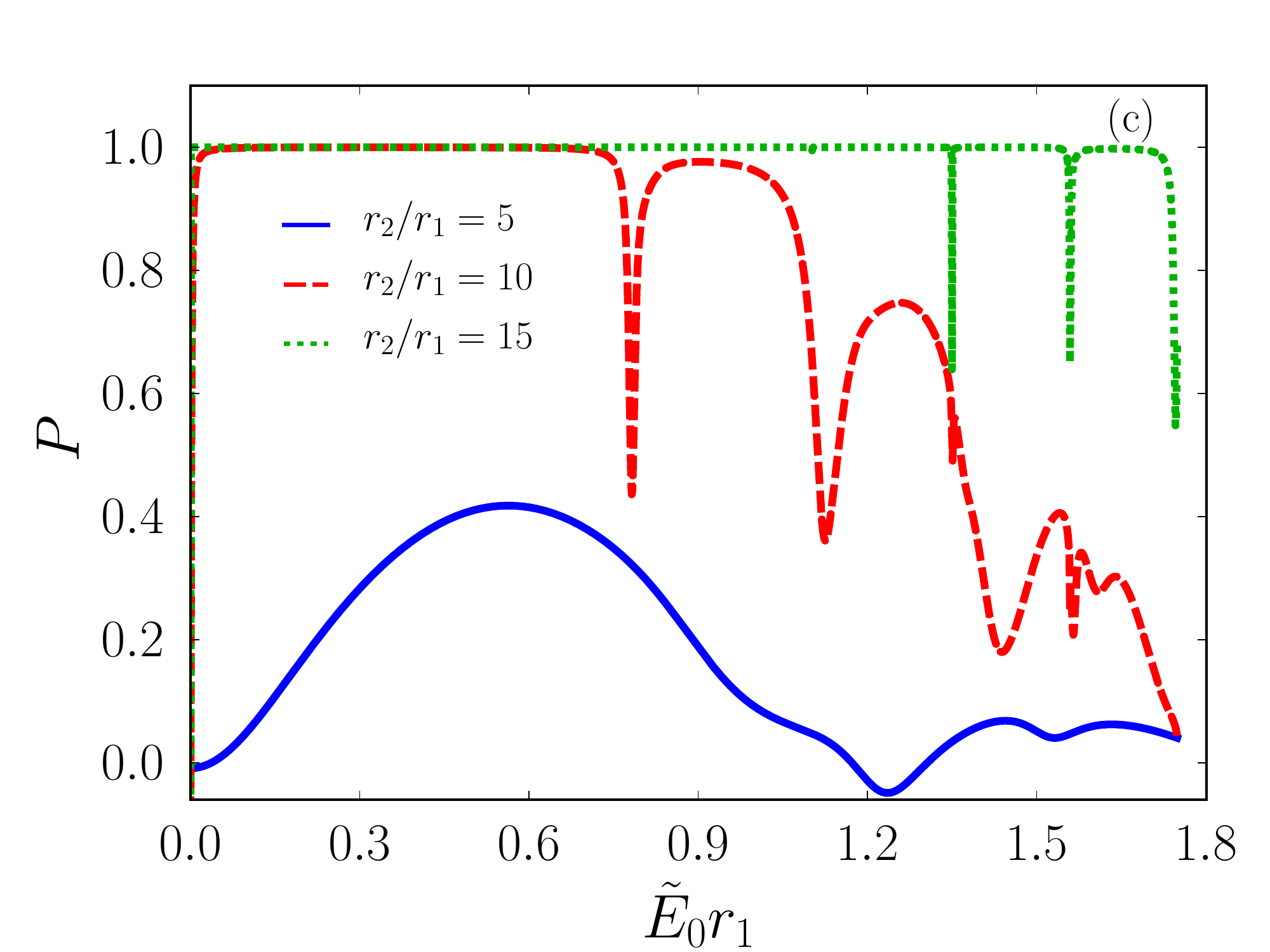}
\caption{(Color online) (a) Conductance as a function of real magnetic flux piercing the middle area for different values of pseudomagnetic flux in zero doped Corbino. The amplitude of the first harmonic oscillation depends on the value of the pseudomagnetic flux and vanishes at $\phi^{el}=\phi_0/4$ due to the destructive interference of electrons at two valleys. (b) Conductance and (c) polarization as a function of doping where the uniaxial strain is zero and an inhomogeneous strain is only applied. (b) indicates the difference between the conductance in the two valleys for the case $\lambda=1$ which causes $B_K=2B$ and $B_{K'}=0$. $B=1T$ and $r_2=10r_1$. Inset: Field suppression disappears due to the zero field strength.}
\label{fig5}
\end{figure}

We showed, in Fig. \ref{fig3}, that the period of this oscillation can be controlled by using uniaxial elastic deformation, while the amplitude of oscillation does not change. We also investigate the effect of the inhomogeneous strain, which creates a constant pseudomagnetic field, on the amplitude of the oscillation. The conductance as a function of real magnetic flux piercing the middle area for different values of pseudomagnetic flux in zero doped Corbino is shown in Fig.~\ref{fig5} (a) which is calculated by Eq.~(\ref{trans0}). It is clear that although at $\phi^{el}=\phi_0/6$, where $\phi_0=2(h/e)\mathrm{ln}(r_2/r_1)$,
the period of the oscillation does not change however the amplitude of the oscillation reduces to its magnitude in the case of zero pseudomagnetic flux. Moreover, for the case of $\phi^{el}=\phi_0/4$ the second harmonic demonstrates a much bigger contribution to the conductance since the first one vanishes. These results can be easily understood considering only the first term of the expansion in Eq.~(\ref{gsdisc}), $G=G^K+G^{K'}\approx2G_0+2G_1\cos(2\pi\phi^{el}/\phi_0)\cos(2\pi\phi/\phi_0)\}$ which means that the first harmonic vanishes at $\phi^{el}=\phi_0/4$ due to the destructive interference of electrons at two valleys.\par

We find, as a result of ${\bf B}^{el}=\tau {\bf B}$ at two valleys and $\epsilon=0$, two different features for the conductance as indicated in Fig.~\ref{fig5}(b). In the absence of the intervalley scattering, electrons in the $K$ point have to obey localized states which are originated from the Landau levels, but at the $K'$-point charges do not feel any magnetic field localization so that total current in the field suppression region at the $K$ point is valley polarized. In other words, the valley filtering is mostly based on the orbital aspect of the wave function where the electrons in two valleys have two different magnetic lengths. In Fig.~\ref{fig5}(c) the valley polarization versus doping rate is shown in different sizes and it is clear that there is a wide energy interval in which $P=1$. This interval increases with increasing Corbino disc size. Note that in the field suppression regime at the $K$ point, the total conductance ($G_K+G_{K'}$) comes from the contribution of the electrons at the $K'$ point which are no longer localized by the magnetic field. This simple picture is practical due to the edgeless structure of the Corbino geometry.

\begin{figure}[!htbp]
\includegraphics[width=\linewidth]{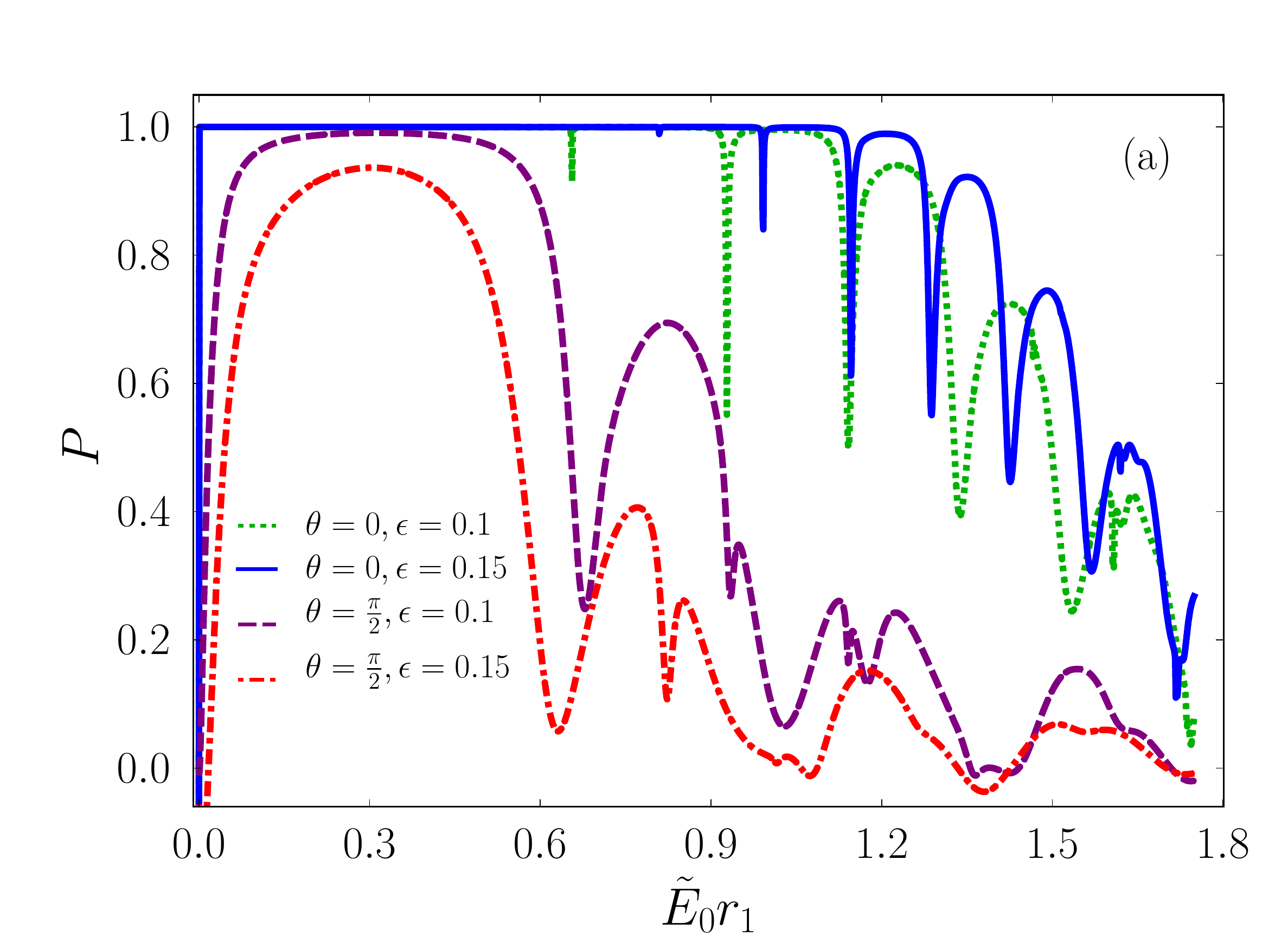}
\includegraphics[width=\linewidth]{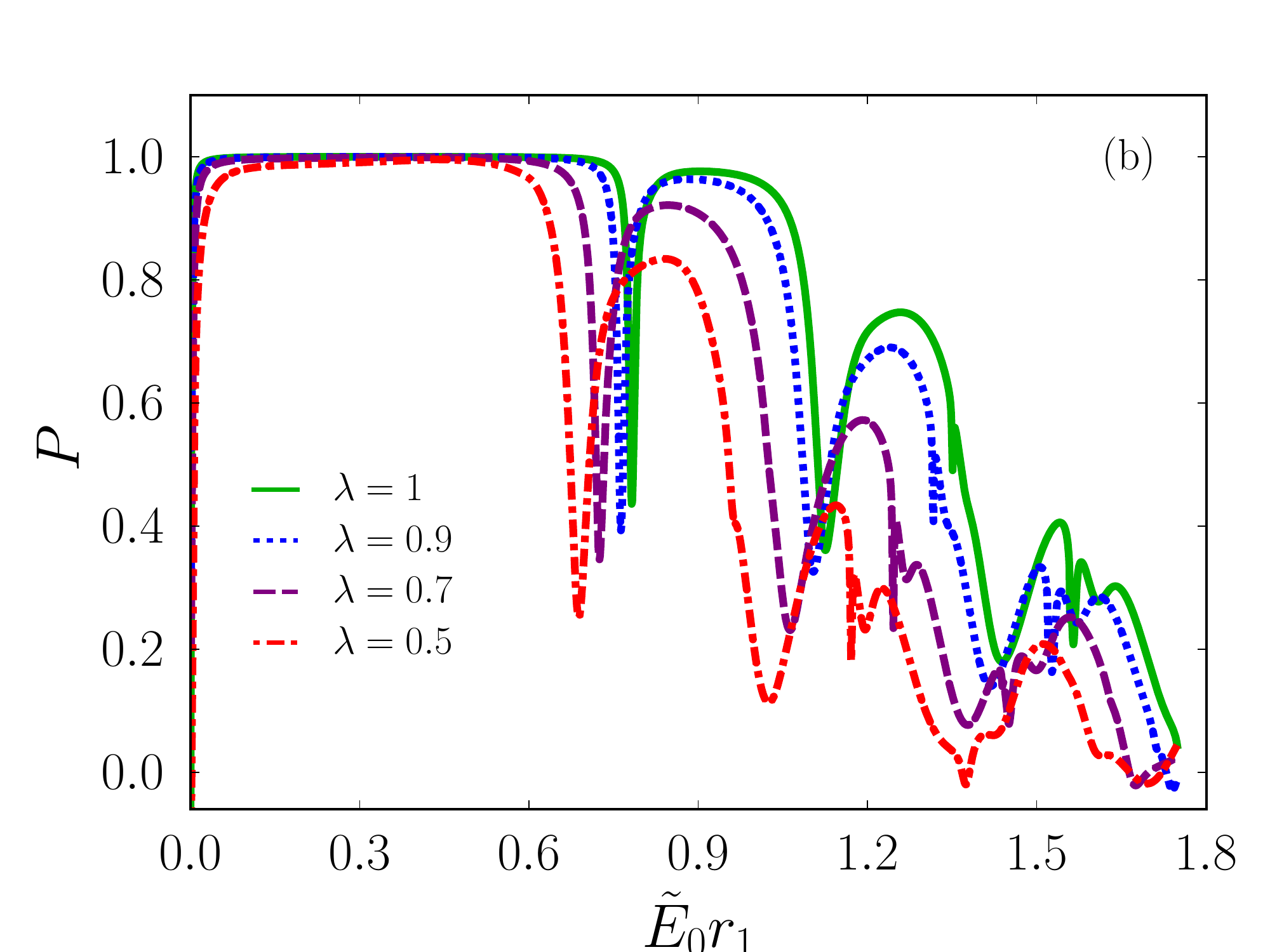}
\caption{(Color online) Valley polarization as a function of the doping for (a) diverse directions and for (b) different $\lambda$ magnitudes. (a) illustrates an anisotropic polarization along the $x$ and the $y$ directions, produced by the uniaxial and inhomogeneous strain.
(b) shows how the polarization persists even in the case in which the pseudomagnetic field is not exactly equal to the real magnetic field. Note that $r_2=10r_1$.}
\label{fig6}
\end{figure}

Moreover, the polarization is tunable by applying a uniaxial strain which is indicated in Fig.~\ref{fig6}(a). Producing the pseudomagnetic field which is exactly equal to the real one, needs a fine tuning process which experimentally would be difficult. We therefore assume the $\lambda$ is not unity and the result is demonstrated in Fig.~\ref{fig6}(b) which shows the persistence of the polarization in this case. As the numerical results show, our physical conclusions regarding the valley polarization are applicable in the inhomogeneous strain.

\section{conclusion}

We have calculated the magnetotransport of a strained graphene in a Corbino geometry. We have shown that, using strains both inhomogeneous and uniaxial in the absence of the edge scattering, the conductance is suppressed in one valley in such a way that the bulk conductance becomes valley polarized in a desired direction, whereas the valleys take part in the conduction in the cross direction. We have investigated the effect of strain on the oscillating nature of the conductance of the system by carrying out an accurate analytic and numeric study. We have found that the oscillating period depends on the value and the sign of the uniaxial strain and also its amplitude can be manipulated by the induced pseudomagnetic flux which  originates from inhomogeneous strain.
By applying a real magnetic field on the strained Corbino system, we have shown that, in the absence of the intervalley scattering, electrons in one valley have to obey localized states which originates from the Landau levels but at the other Dirac point charges do not feel any magnetic field localization therefore total current is valley polarized in the field suppression region at one valley. Furthermore, we have obtained the valley polarization by applying an inhomogeneous and uniaxial strain on the Corbino disc and its dependence on the size, the value of the uniaxial strain and also the directions of the Corbino deformation. The size dependence of the system has a dramatic effect on the conductance and we have shown that stronger field suppression appears in a larger Corbino which can be helpful to generate the valley polarized current. Our analysis can be generalized for a bilayer graphene in a Corbino geometry.

\begin{acknowledgments}
Useful discussions with A. Concha in the early stage of this work are acknowledged.
\end{acknowledgments}


\begin{thebibliography}{99}

\bibitem{geim}
A. K. Geim, K. S. Novoselov, Nature Matherial {\bf 6}, 183 (2007).

\bibitem{vozmediano}
M. A. H. Vozmediano, M. I. Katsnelson, F. Guinea, Phys. Reports {\bf 496}, 109 (2010).

\bibitem{Goerbig08}
M. Goerbig, J.-N. Fuchs, G. Montambaux, F. Pi\'{e}chon, Phys. Rev. B {\bf 78}, 045415 (2008).

\bibitem{Pereira09}
V. M. Pereira, A. H. C. Neto, and N. M. R. Peres, Phys. Rev. B {\bf 80}, 045401 (2009); V. M. Pereira, R. M. Ribeiro, N. M. R. Peres, A. H. Castro Neto, Eur. Phys. Lett. {\bf 92}, 67001 (2010).

\bibitem{Ribeiro09}
R. M. Ribeiro, V. M. Pereira, N. M. R. Peres, P. R. Briddon, and A. H. C. Neto, New J.
Phys. {\bf 11}, 115002 (2009).

\bibitem{guinea1}
F. Guinea, M. I. Katsnelson, A. K. Geim, Nature Physics {\bf 6}, 30 (2010); F. Guinea, A. K. Geim, M. I. Katsnelson, and
K. S. Novoselov, Phys. Rev. B {\bf 81}, 035408 (2010).


\bibitem{diana}
D. A. Gradinar, M. Mucha-Kruczyn'ski, H. Schomerus, and V. I. Fal'ko
Phys. Rev. Lett. {\bf 110}, 266801 (2013).

\bibitem{Rostami}
H. Rostami and R. Asgari, Phys. Rev. B \textbf{86}, 155435 (2012).

\bibitem{suzuura}
H. Suzuura and T. Ando, Phys. Rev. B {\bf 65}, 235412 (2002).

\bibitem{maes}
J. L. Ma\`{e}s, Phys. Rev. B {\bf 76}, 045430 (2007).

\bibitem{lee08}
C. Lee, X. Wei, J. W. Kysar, J. Hone, Science {\bf 321}, 385
(2008); W. Bao, F. Miao, Z. Chen, H. Zhang, W. Jang, C. Darmes, C.
N. Lau, Nat. Nanotechnology {\bf 4}, 562 (2009).

\bibitem{Levy10}
N. Levy, S. A. Burke, K. L. Meaker, M. Panlasigui, A. Zettl1, F.
Guinea, A. H. Castro Neto and M. F. Crommie, Science {\bf 329 }, 554 (2010).

\bibitem{nima}
N. Abdedpour, R. Asgari and F. Guinea, Phys. Rev. B {\bf 84},
115437 (2011).

\bibitem{fogler}
M. M. Fogler, F. Guinea, and M. I. Katsnelson, Phys. Rev. Lett.
{\bf 101}, 226804 (2008).

\bibitem{bunch}
J. S. Bunch, S. S. Verbridge, J. S. Alden, A. M. van der Zande, J.
M. Parpia, H. G. Craighead, and P. L. McEuen, Nano Lett. {\bf 8},
2458 (2008).

\bibitem{exp}
W. Bao, F. Miao, Z. Chen, H. Zhang, W. Jang, C. Dames,
and C. N. Lau, Nat. Nanotechnol. {\bf 4}, 562 (2009); M.L.
Teague, A. P. Lai, J. Velasco, C. R. Hughes, A. D. Beyer,
M.W. Bockrath, C. N. Lau, and N.-C. Yeh, Nano Lett. {\bf 9},
2542 (2009);
Keun Soo Kim, Yue Zhao, Houk Jang, Sang Yoon Lee, Jong Min Kim, Kwang S. Kim, Jong-Hyun Ahn,
Philip Kim, Jae-Young Choi, and Byung Hee Hong, Nature (London) {\bf 457}, 706 (2009);
T. M. G. Mohiuddin, A. Lombardo, R. R. Nair, A. Bonetti, G. Savini, R. Jalil, N. Bonini,
D. M. Basko, C. Galiotis, N. Marzari, K. S. Novoselov, A. K. Geim, and A. C. Ferrari, Phys. Rev. B {\bf 79}, 205433 (2009);
Changgu Lee, Xiaoding Wei, Jeffrey W. Kysar, and James Hone, Science {\bf 321}, 385 (2008); Mingyuan Huang, Hugen Yan, Tony F. Heinz, and James Hone
Nano Lett. {\bf 10}, 4074 (2010).

\bibitem{guinea}
F. Guinea, Baruch Horovitz, and P. Le Doussal
Phys. Rev. B {\bf 77}, 205421 (2008).

\bibitem{low}
Tony Low, F. Guinea, Nano Lett. {\bf 10}, 3551 (2010).

\bibitem{boltzmann}
L. Boltzmann, Philos. Mag. {\bf 22}, 226 (1886).

\bibitem{Galdamini}
S. Galdamini and G. Giuliani, Ann. Sci., {\bf 48}, 21 (1991).

\bibitem{Rycroft}
 S. F. W. R. Rycroft, R. A. Doyle, D. T. Fuchs, E. Zeldov,
R. J. Drost, P. H. Kes, T. Tamegai, S. Ooi, D. T. Foord, Phys. Rev. B \textbf{60}, 757(R) (1999).

\bibitem{Yan10}
J. Yan and M. S. Fuhrer, Nano Lett. \textbf{10} (11), 4521 (2010).

\bibitem{Faugeras10}
C. Faugeras, B. Faugeras, M. Orlita, M. Potemski, R. R. Nair, and A. K. Geim, ACS Nano \textbf{4}(4), 1889 (2010).

\bibitem{zhao12}
Y. Zhao, P. C.-Zimansky, F. Ghahari, and P. Kim, Phys. Rev. Lett. \textbf{108},106804 (2012).

\bibitem{Wiegers}
S. A. J. Wiegers ,J. G. S. Lok, M. Jeuken ,U. Zeitler ,J. C. Maan ,M. Henini,Phys. Rev. B \textbf{59}, 7323�7326 (1999).


\bibitem{Rycerz09}
A. Rycerz, P. Recher, and M.Wimmer, Phys. Rev. B {\bf 80}, 125417 (2009).

\bibitem{Nazarov}
Y. V. Nazarov, Y. M. Blanter, \emph{Quantum Transport: Introduction to Nanoscience}, Cambridge University Press, Cambridge (2009).

\bibitem{Rycerz10}
A. Rycerz, Phys. Rev. B {\bf 81}, 121404R (2010).

\bibitem{Katsnelson10}
M. I. Katsnelson, Europhys. Lett.  \textbf{89} 17001 (2010).

\bibitem{Rycerz12}
Adam Rycerz, Acta Phys. Polon. A \textbf{121}, 1242 (2012).

\bibitem{Shkolnikov02}
Y. P. Shkolnikov, E. P. De Poortere, E. Tutuc , and M. Shayegan, Phys. Rev. Lett. \textbf{89}, 226805 (2002).

\bibitem{Beenakker}
A. Rycerz, J. Tworzydo and C. W. J. Beenakker, Nature Physics \textbf{3}, 172 (2007).

\bibitem{Xiao07}
D. Xiao, W. Yao, and Q. Niu, Phys. Rev. Lett. \textbf{99}, 236809 (2007).

\bibitem{Isberg13}
J. Isberg, M. Gabrysch, J. Hammersberg, S. Majdi, K. K. Kovi, and D. J. Twitchen,Nature Materials \textbf{12}, 760 (2013).

\bibitem{Zhai10}
F. Zhai, X. Zhao, K. Chang, H. Q. Xu, Phys. Rev. B {\bf 82}, 115442 (2010).

\bibitem{Chaves10}
A. Chaves, L. Covaci, Kh. Yu. Rakhimov, G. A. Farias, F. M. Peeters, Phys. Rev. B {\bf 82}, 205430 (2010).

\bibitem{Jiang13}
Y. Jiang, T. Low, K. Chang, M. I. Katsnelson, F. Guinea, Phys. Rev. Let {\bf 110}, 046601 (2013).

\bibitem{Neto07}
A. H. Castro Neto and F. Guinea, Phys. Rev. B \textbf{75}, 045404 (2007).

\bibitem{Gunlycke11}
D. Gunlycke and C.T. White, Phys. Rev. Let, {\bf 106}, 136806 (2011).

\bibitem{Schomerus10}
H. Schomerus, Phys. Rev. B, {\bf 82}, 165409 (2010).

\bibitem{PereiraJr09}
J.M. Pereira Jr, F. M. Peeters, R. N. Costa Filho, G. A. Farias, Journal of Physics: Condensed Matter, {\bf 21}, 045301 (2009).

\bibitem{Klimov12}
N.N. Klimov, S. Jung, S. Zhu, T. Li, C. Wright, S. D. Solares, D. B. Newell, N. B. Zhitenev, J. A. Stroscio, Science, {\bf 336}, 1557 (2012).

\bibitem{Fernando13}
F. de Juan, J. L. M\'{a}nes, and M.A. H. Vozmediano, Phys. Rev. B \textbf{87}, 165131 (2013).

\bibitem{mathfunc}
M. Abramowitz and I. A. Stegun \textit{Handbook of Mathematical Functions} (Dover, New York, 1965).

\bibitem{beenakker}
C. W. J. Beenakker, Rev. Mod. Phys. {\bf 80}, 1337 (2008).

\bibitem{Prada07}
E. Prada, P. San-Jose, B. Wunsch, F. Guinea, Phys. Rev. B {\bf 75}, 113407 (2007).

\bibitem{Zhai12}
F. Zhai, K. Chang, Phys. Rev. B {\bf 85}, 155415 (2012).

\end{thebibliography}
\end{document}